# Report on a pre-earthquake signal detection by enhanced Eötvös torsion balance

L. Völgyesi [1], Gy. Tóth [1], Gy. Szondy [2], B. Kiss [3], E. Fenyvesi [4], G.G. Barnaföldi [4], Cs. Égető [1], P. Lévai [4], E. Imre [5], M. Pszota [4,6], P. Ván [4,7*]

[1] Department of Geodesy and Surveying, Faculty of Civil Engineering, Budapest University of Technology and Economics, Műegyetem rkp. 3., H-1111 Budapest, Hungary
[2] Independent researcher, Society for the Unity of Science and Technology, Hungary
[3] Department of Control Engineering and Information Technology, Faculty of Electrical Engineering and Informatics, Budapest University of Technology and Economics, Műegyetem rkp. 3., H-1111 Budapest, Hungary
[4] Department of Theoretical Physics, Institute for Particle and Nuclear Physics, Wigner Research Centre for Physics, Konkoly-Thege Miklós út 29-33., H-1121 Budapest, Hungary
[5] Óbuda University, Faculty of Electrical Engineering, Bécsi út 94-96 BC I.107, H-1034 Budapest, Hungary
[6] Eötvös Loránd University, Faculty of Science, Pázmány Péter sétány 1/A, H-1117 Budapest, Hungary
[7] Department of Energy Engineering, Faculty of Mechanical Engineering, Budapest University of Technology and Economics, Műegyetem rkp. 3. H-1111 Budapest, Hungary
*email: van.peter@wigner.hu

**Abstract**

More than 30 minutes before the earthquake event in Florina, Greece at 21:43:47(UTC) on 9$^{th}$ January 2022 an enhanced Eötvös torsion balance registered unidentified signals in the Jánossy Underground Research Laboratory in Budapest, Hungary. These signals were not visible on the seismograms, although seismic noises are continuously recorded by a side-by-side broadband seismometer. Moreover, seismological stations did not detect anything unusual, they presented a negative confirmation of the events. Our observation suggests that torsion-balances might effectively detect precursory earthquake signals from a considerable distance. Such a finding could trigger the development of new observational devices and networks and can provide novel knowledge about the origin and mechanism of earthquake phenomena.

**Keywords**: Eötvös torsion balance, microseismical noise, equivalence experiment, earthquake prediction

# Introduction

Earthquake prediction is an important goal of earth science research. The primary source of data is the worldwide network of seismological stations. However, it is not excluded that different and more sensitive instruments can detect signals of earthquake-related geophysical events that traditional seismometers cannot observe. Therefore, a wide variety of parameters are monitored to determine earthquake risks, including unusual changes in the altitudes, tilts, occurrence of micro-earthquakes and electromagnetic signals[1-6]. Despite the efforts to achieve reliable forecasting, exact short-time earthquake prediction is still not feasible.

Our work aims to report a recent measurement of a signal detected by an enchanced Eötvös torsion balance prior to the Florina earthquake, Greece.

# Results

a) **The instruments**. The signal prior to the earthquake was detected by an *Eötvös type gravity gradiometer* with a sensitivity of 0.01 Eötvös with data acquisition of 1 Hz. The in-strument is an improved original small Eötvös-Pekár torsion balance, originally developed for in situ geophysical measurements[7,8]. The original sensitivity of the device is 1 Eötvös, and the two-magnitude improvement is due to a combination of several hardware and software developments, like camera readout, automatic rotation, precautious code ring positioning and control, AI-based program package of image recognition and torsion angle determination with Fourier analysis, program package for determining the equilibrium from the dynamics of the balance using the whole time series. Further details of the modifications and developments are given as Supplementary material.

The earthquake was detected also by a *broadband Güralp 3T three-axis seismometer*, installed next to the balance at 1 m distance.

b) ***The site*** of our experiment is the Jánossy Underground Research Laboratory (JURLab) unit of the Vesztergombi Laboratory of High Energy Physics (VLAB) at Wigner Research Center for Physics. The laboratory is outside the city and the observation chamber, where the instruments are installed, is at a depth of 30 meters below ground. On Fig. 1 the Eötvös-Pekár torsion balance and the Güralp 3T three-axis broadband seismograph are shown in the laboratory. The temperature stability of this site is satisfactory, the daily change is only a few hundredths of a degree and the amplitude of the microseismical ground vibrations is also sufficiently small. Environmental factors are continuously recorded with several temperature, pressure, humidity and occasionally by infrasound sensors.

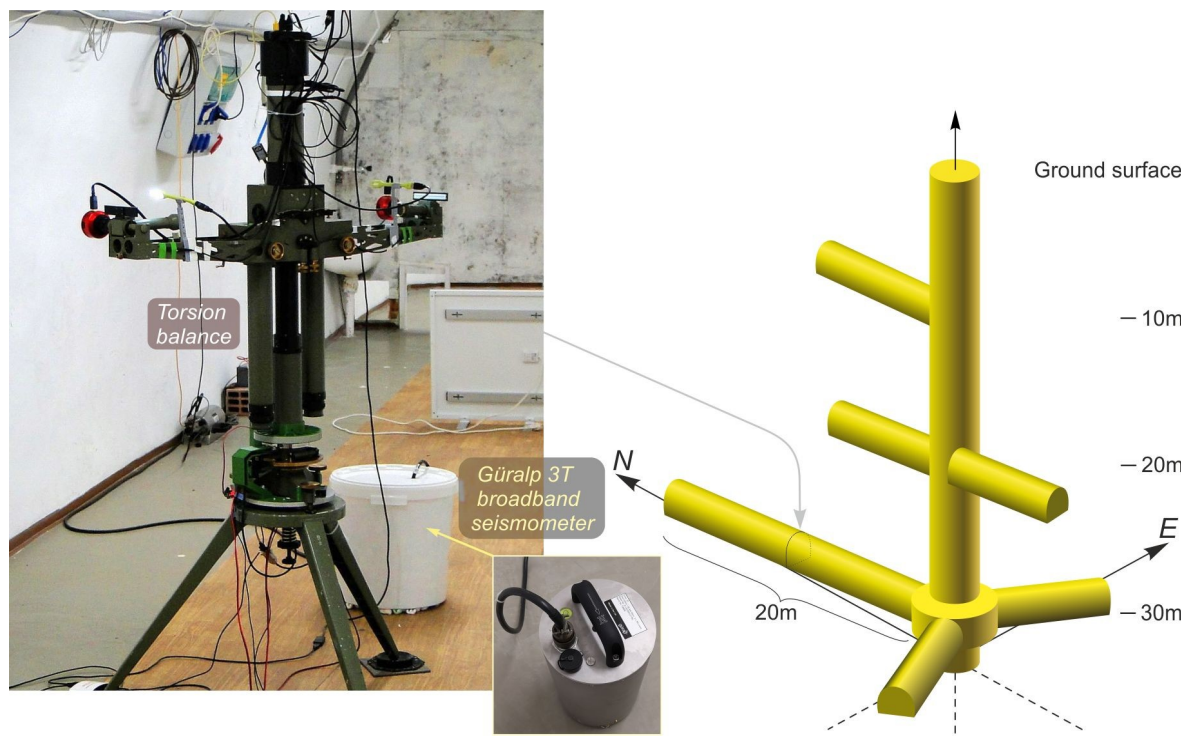

***Fig. 1***. *The location of Eötvös-Pekár torsion balance and the Güralp 3T seismometer at the Jánossy Underground Research Laboratory (JURLab).*

Our torsion balance is an excellent tool for recording earthquake waves. In the last 2 years, we have recorded around 60 earthquakes of greater or lesser magnitude.

In December 2021 the dynamical properties of the second pendulum of the double balance were modified, because of the installation of an Al test mass. Therefore, in our double balance, pendulum I has an Au-Au mass pair and pendulum II has an Au-Al one. Then an earthquake was recorded, and our torsion balance registered unusual vibrations approximately half an hour before the earthquake. Furthermore, the Güralp 3T seismograph, located 1 m from the Eötvös torsion balance did not register any unusual activity prior to the earthquake, a normal signal was recorded at the official arrival time.

This earthquake occurred in Florina, Greece on 9 January 2022, (the parameters: $t_o$ = 21:43:47/UTC/, 40°43'48.0"N, 21°21'36.0"E, Richter magnitude $M$ = 5.6, focal depth $h$ = 9km). Seismograms of this earthquake were registered among others at the Kövesligethy Radó Seismological Observatory in Budapest (47.48°N, 19.02°"E) see Fig.2 [*WEB*]. This is a regular seismogram, without any special property.

In contrast to the usual seismograph recordings, the Eötvös torsion balance shows unusual and peculiar vibrations. The earthquake pushed the pendulum II (red color) of the torsion balance out of equilibrium and then disturbance was damped, and the equilibrium was restored (top of the Fig. 3). However, with a higher resolution one can see that a characteristic signal appeared about 32 minutes before the earthquake (in the middle part of Fig. 3). The lower part of Fig. 3 shows a highly magnified picture of the shaking of pendulum II in the half hour before the earthquake.

The behaviour of pendulum I (blue colour) of the torsion balance is also interesting, which is fundamentally different from that of pendulum II. It can be seen on Fig. 5 that pendulum I does not show the special signal seen on pendulum II, even when magnified. However, Fig. 6. clearly shows that pendulum I (blue) with a little delay compared to pendulum II (red) also detected something, and more than half an hour before the earthquake's origin, the direction of deflection of pendulum I has abruptly changed.

It is very important to note that prior to the earthquake, the Güralp 3T seismograph installed in the immediate vicinity of the Eötvös torsion balance did not indicate any unusual waves. This can be clearly seen in Fig. 6, which shows the vertical component of the ground motions. Likewise, the seismograms of the two horizontal components do not show any event before the arrival of the earthquake waves.

We did not register any similar signals before this observation. There were not any activities inside the laboratory, nor in the vicinity, in this period. Neither the thermometers, the barometers nor the neighbouring seismograph indicated any external disturbances that could be correlated to the unusual signal. The torsion balance was all the time at rest, it was not rotated.

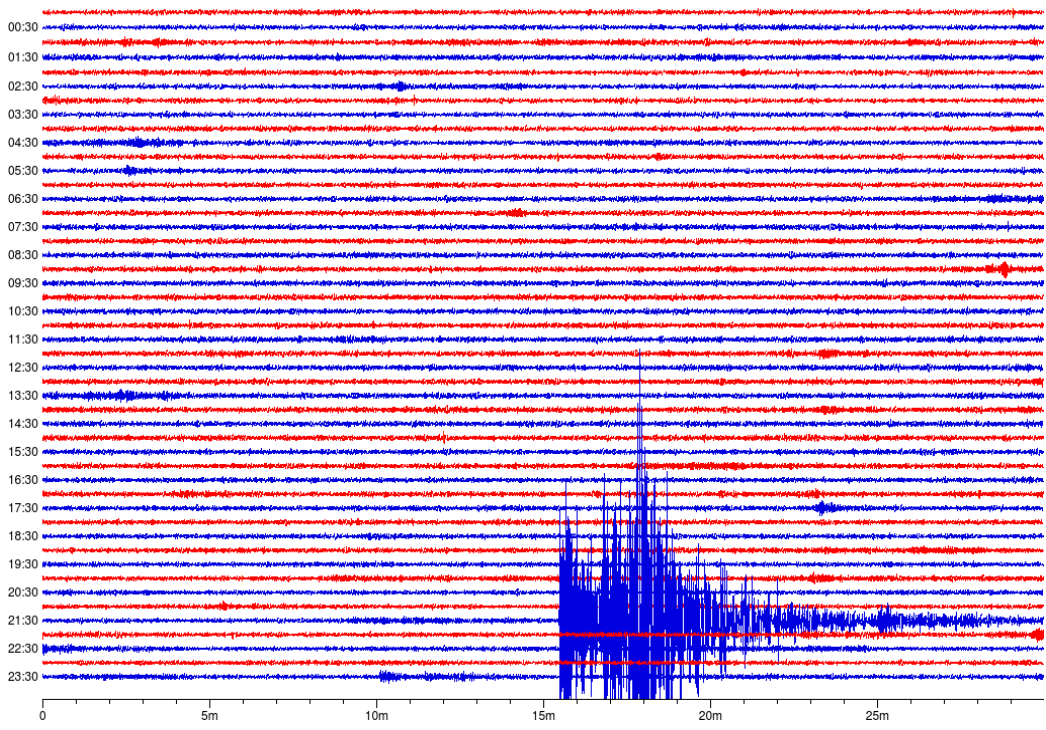

*Fig. 2*. *Seismogram of the Florina earthquake in Budapest. Time is in UTC* [*WEB*].

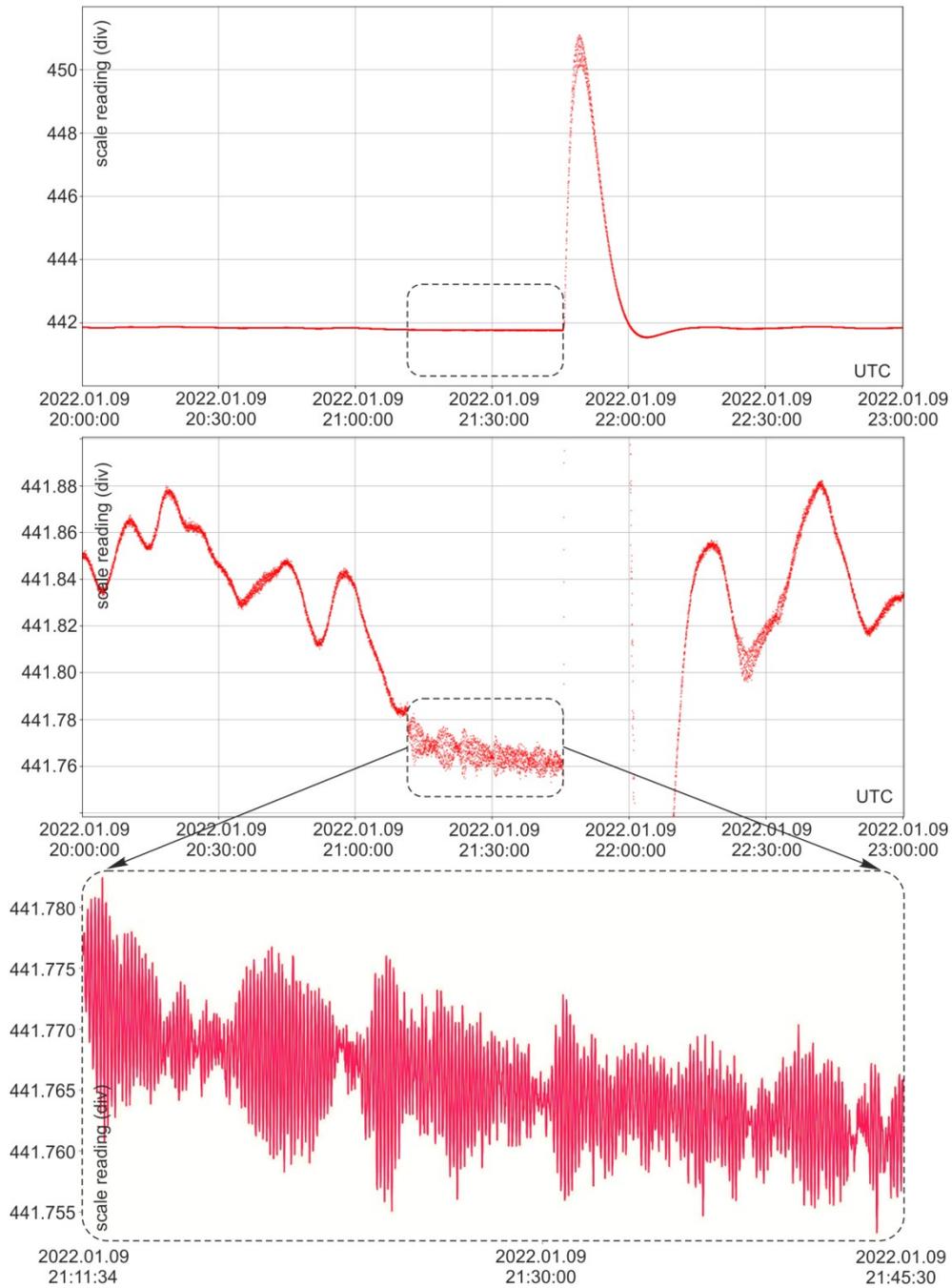

***Fig. 3.*** *Earthquake in Florina, Greece on 9 January 2022, registration and enlarged images by pendulum II of the torsion balance.*

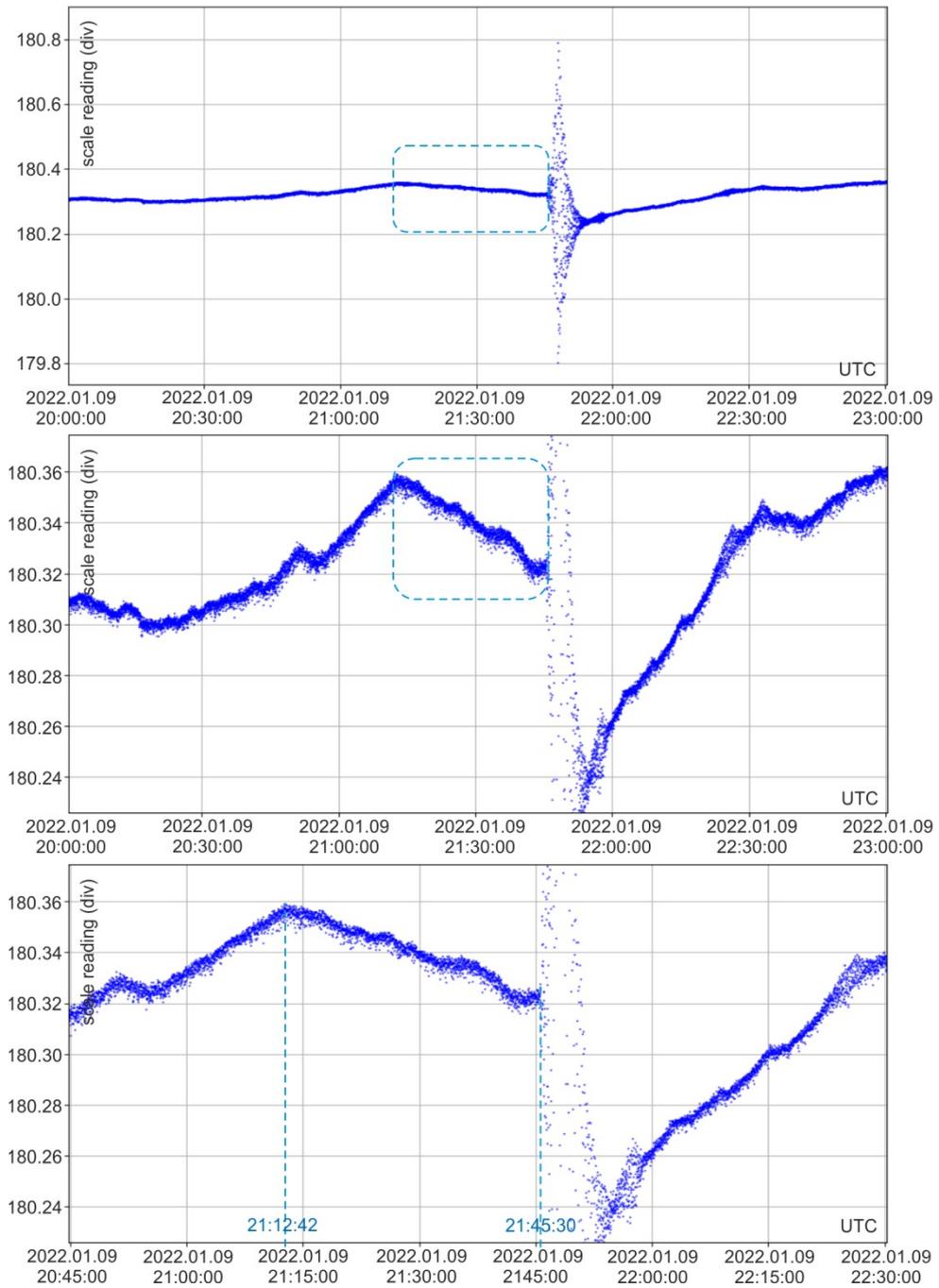

***Fig. 3****. Earthquake in Florina, Greece on 9 January 2022, registration and enlarged images by pendulum I of the torsion balance.*

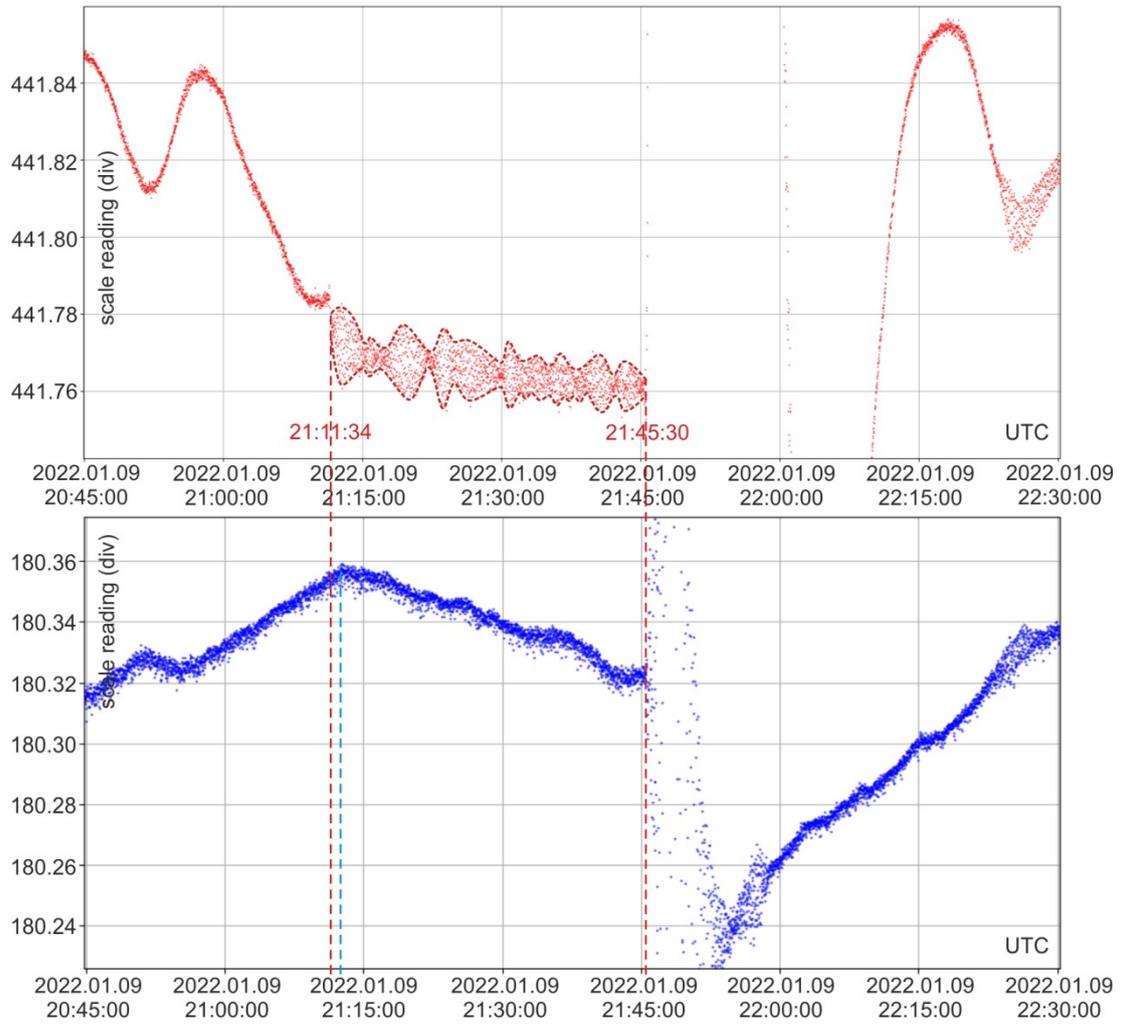

***Fig. 4***. *Comparing the registration of pendulums II and I.*

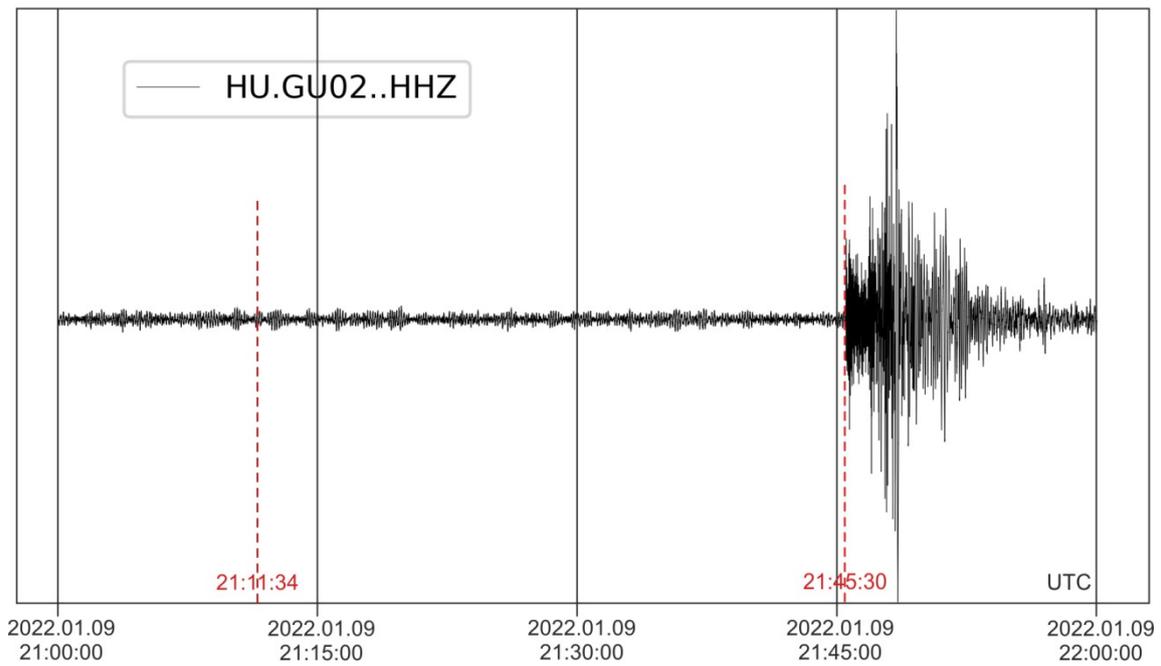

***Fig. 5***. *Registration of the vertical component of the Greek earthquake by the Güralp 3T seismometer next to the torsion balance*

## Discussion

Eötvös balance is a unique device that is sensitive to the force difference of its test masses with extreme sensitivity. The lack of seismometer activity excludes the possibility of coincidental mechanical vibrations of human origin as an explanation. Also, the environmental records exclude the possibility of thermal and pressure related effects as possible explanations of the observed pre-earthquake signal. With the recent sensitivity of 0.01 Eötvös, our Eötvös-type gravity gradiometer can detect an extra point-like 100 kg mass in about 56 m from the balance. Considering also that the characteristic torsional response time of the instrument is 1226 s, direct gravitational effect can be excluded, only extraordinary changes of the mass distribution could produce a signal that is similar to the observed one.

*Therefore, the lack of observable environmental effects and the timing may indicate that the signal is not coincidental but can be related to some pre-earthquake phenomena.*

A possible explanation of our observation can be related to changes in the spectral distribution of microseismical noise. Eötvös-type gravity gradiometers are extremely sensitive frequency-selective amplifiers near to their resonance frequencies due to their characteristic dipole construction and suspension system[9,10]. This feature of the balance is available only with continuous registration of the position and with high sensitivity. The characteristic properties of our torsion balance based spectral analyzer are $1$-$10^{-3}$ Hz bandwidth and $10^{-11}$ m sensitivity. That is far better than spectral analyses based on seismometer data (about $10^{-2}$ Hz and $10^{-8}$ m). Therefore, one cannot detect our signal with the recent technology of seismological observations. Similar, Eötvös-type gravity gradiometers were developed in the Institute of Experimental Geophysics, Russian Federation, motivated by the possibility of earthquake prediction[11-13] and the theoretically predicted possibility of amplification was experimentally confirmed. Our observed signal differs from the signal of the previously reported observations, but the structure of the modernized Eötvös balance is also different, except for the dipole construction.

Further studies are needed to clarify and understand the above hypothesis. Also, the mechanism of spectral changes of the microseismical noise prior to the earthquake requires explanation. Further, independent detection of similar signals is necessary for the clarification of the phenomena.

## Acknowledgements

The work was supported by the grants National Research, Development and Innovation Office NKFIH 123815. The support of TKP is acknowledged. This work was performed in the frame of the FIEK 16-1-2016-0007 project, implemented with the support provided from the National Research, Development and Innovation Fund of Hungary, financed under the FIEK 16 funding scheme. Some parts of the research reported in this paper have been supported by the National Research, Development and Innovation Fund (TUDFO/51757/2019-ITM), Thematic Excellence Program. Special thanks to the Vesztergombi Laboratory of High Energy Physics (VLAB), funded by the Wigner Research Centre for Physics, for the continuous technical support in the Jánossy Underground Research Laboratory JURLab).

## Data availability statement

The data that support the findings of this study are available from the corresponding author upon reasonable request.

## Methods

**Eötvös type torsion balances**

**a. Historical notes**

Between 1906 and 1908, *Loránd Eötvös* and his colleagues carried out their famous equivalence experiment (known as *Eötvös-Pekár-Fekete* (EPF) measurement) for validating the equivalence of gravitational and inertial mass[7]. The original measurements contain tiny regular differences, the explanation of which is a matter of debate among physicists[14-16]. In 2017, we decided to repeat the experiment under modern technical conditions[8]. After long preparations, we started our measurements in 2019[17,18], which are still ongoing continuously.

The basic instrument of these measurements is the best-known gravity gradiometer, the Eötvös torsion balance. The torsion balance consists of a horizontal beam having the length $2l$ with masses $m$ on each end, suspended with a torsion wire. One of the two masses is on one end of the horizontal beam, and the other mass is suspended at a distance $h$ from the other end of the beam – as can be seen on Fig. s1. The horizontal component of gravity acting on the two masses causes a torque, and the horizontal beam is rotated until an equilibrium position with the restoring torque of the suspending torsion wire (having the torsion constant $\tau$) is reached. In the equilibrium condition of torques the scale reading is $n$, while the scale reading of the unknown torsion-free zero position of the beam would be $n_0$ [19].

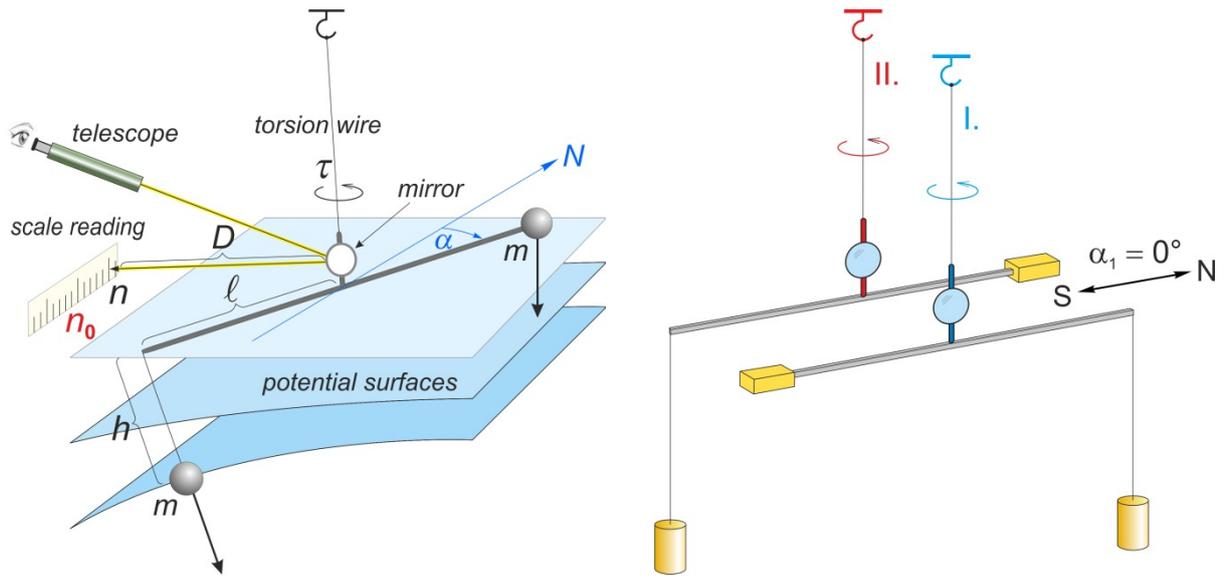

***Fig. s1***. The base principle of the Eötvös torsion balance.

The base equation of the Eötvös torsion balance is:

$$n - n_0 = \frac{DK}{\tau}\left(W_\Delta \sin 2\alpha + 2 W_{xy} \cos 2\alpha\right) + \frac{2Dhlm}{\tau}\left(W_{zy} \cos\alpha - W_{zx} \sin\alpha\right), \qquad (1)$$

where $W_{zx}$ and $W_{zy}$ are the horizontal gradients of gravity, $W_\Delta$ and $W_{xy}$ are the curvature gradients, $K$ is the moment of inertia, $\alpha$ is the azimuth of the beam and $D$ is the optical distance (see on Fig. s1). The earlier type of instrument is the *Cavendish* torsion balance, in which the two masses are on the same height on the two ends of the beam. This type of instrument is unable to measure the components of *horizontal gradient* $W_{zx}$ and $W_{zy}$, because $h = 0$ in Eq. (1).

Based on Eq. (1) there are five unknowns (the scale reading of the torsion-free zero position $n_0$, the horizontal gradients $W_{zx}$, $W_{zy}$ and the curvature gradients $W_\Delta$, $2W_{xy}$) at each measuring site, so the readings should be made in five different azimuths. Usually, two beam systems are mounted in one instrument at an antiparallel position to each other, so there is a new unknown torsion-free zero position $n_0'$ for the other beam system. Due to the additional unknown, a minimum of six measurements in three different azimuths (e.g., 0o, 120o, 240o) are sufficient to determine the Eötvös tensor components, but it is necessary to repeat the measurements in order to increase the accuracy.

### b. Preparations and modernization

The motivation of the modernization was to repeat the EPF experiment in better circumstances, with at least one or two order higher precision using the original method of Eötvös. For the experiment we had three different types of torsion balances (Eötvös-Pekár, Auterbal and E54), see Fig. s2. From these three types, the Eötvös-Pekár balance has seemed to be the best for our experiment because of its high precision and ease of use.

The Eötvös-Pekár torsion balance, which is known as the Small Original Eötvös G-2, has three variants; they differ from each other in the length of the torsion wires only[20] (Szabó 2016). Our model which was manufactured in 1930 has 30 cm long torsion wires. The developer of the instrument, Dezső Pekár, who had long experience in field measurements, aimed for high accuracy and simplicity of the instrument, so he preferred manual rotation and visual readings instead of automatic operation of the Auterbal and E54 instruments. The only disadvantage of the original Pekár balance is the need for the constant presence of the observer. Namely, the gravitational force of the mass of the observer's body changes the damped position of the torsion balance, moreover, going toward the instrument the noise of his steps causes ground vibrations, which also reduce the reading accuracy. We have achieved solutions for these problems with two important enhancements: by applying a computer-controlled scan on a CCD sensor instead of visual reading and by using computer remote-controlled rotation mechanics to rotate the torsion balance into various azimuths.

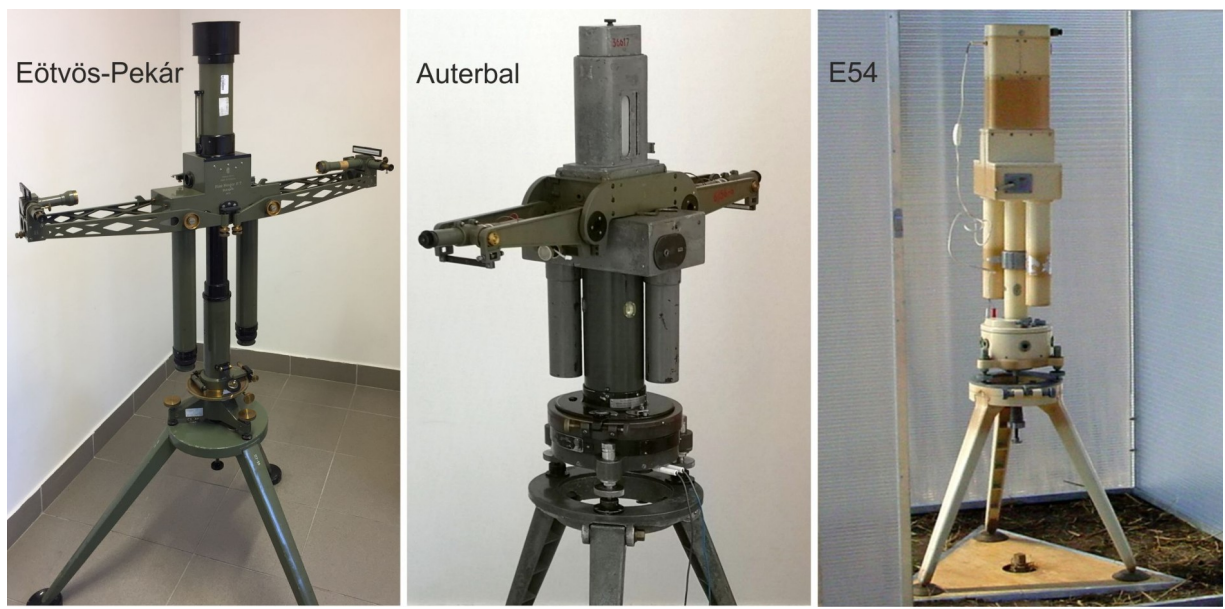

***Fig. s2***. *Three types of instruments: Eötvös-Pekár, Auterbal, and E54 torsion balances.*

By mounting a CCD sensor instead of visual reading (see Fig. s3.), we can determine the very precise damped position of the balance arm by evaluating the resulting digital image. The original optical results in a readout of the equilibrium position with an accuracy of 0.1 scale division, but now with CCD sensors and continuous reading (up to 10 readings/sec), an accuracy of 0.003 scale division is possible. So now our digital reading accuracy is nearly two orders of magnitude better than the reading accuracy of the original Eötvös torsion balance.

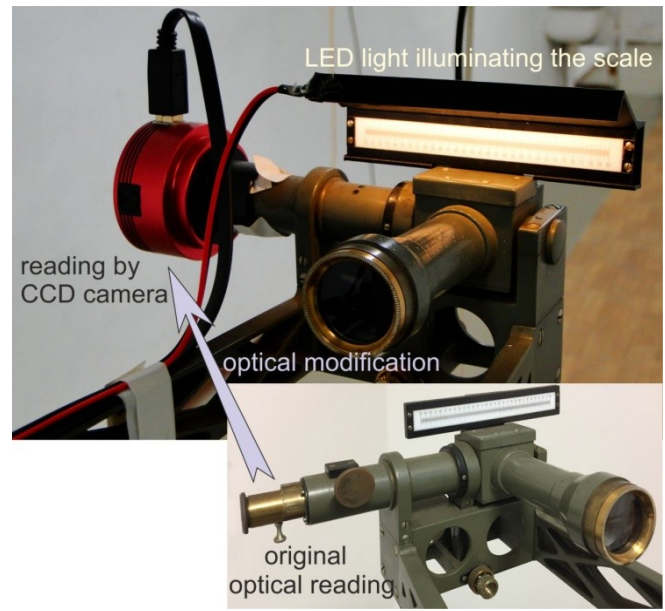

***Fig. s3***. *Replacing the visual reading by CCD camera*

An essential requirement for the experiment is the precise determination of the damped position of the balance arm. In the original EPF experiment, this equilibrium position was determined after a waiting time of 40 to 60 minutes by a single visual reading. Unlike the original experiment, the equilibrium position is determined now by fitting a theoretical damping curve to the real damping curves of the torsion balance arms. Therefore, the damped position can be determined with higher accuracy on the basis of more data, and on the other hand, by continuously monitoring the oscillation of balance bars, we can get important information about the behaviour of the instrument during measurements.

The most critical mechatronic element is the construction of a unit for remote rotation of the instrument. For rotation of the torsion balance into various measuring azimuths, the structure shown in Fig. s4 was made, applying a special motor, which is connected to the upper part of the free rotating torsion balance via a ribbed belt drive.

The rotating mechanism has low weight and low metallic content. We used 3D printing technology (using PLA filament) to produce most of the parts. A low power DC motor rotates the torsion balance through a timing belt with a 1:10 gear ratio hence its magnetic field has negligible influence. A cascade position control loop is realized on the microcontroller with a 100 Hz sampling frequency which meets the desired closed-loop requirements. Since the precise reading of the azimuth value is an important requirement for the overall measurement accuracy, a high resolution absolute optical encoder is used with an accuracy of under 3" (arcsec).

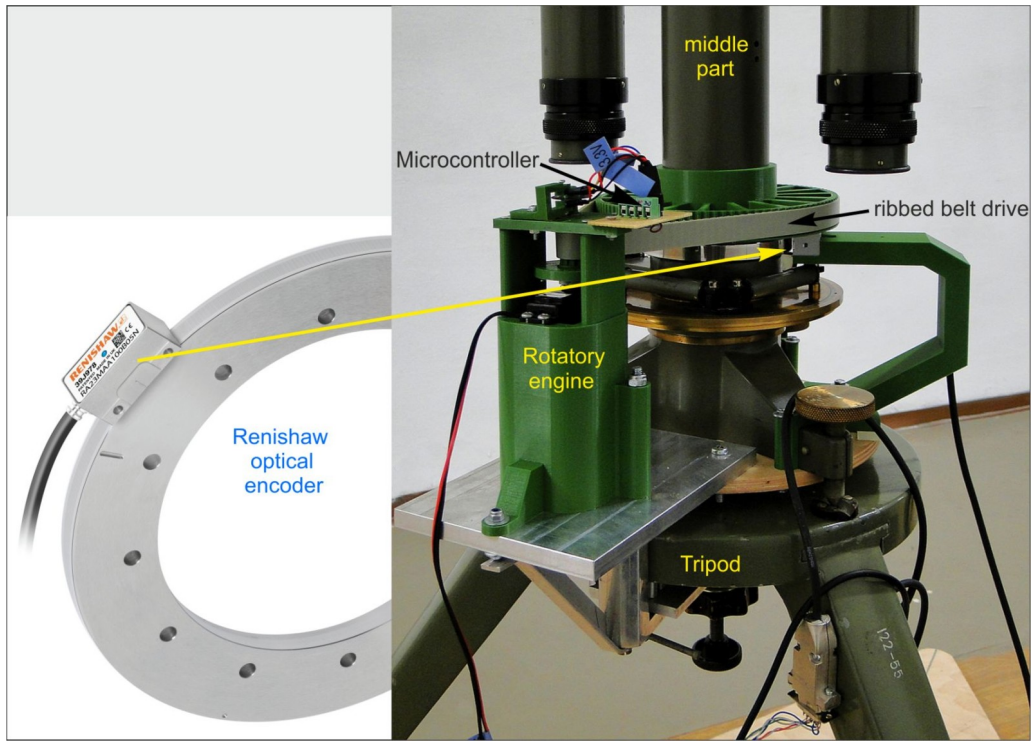

***Fig. s4***. *The remote controlled rotation mechanism of the torsion balance*

The architecture of the data acquisition and control system is depicted in Fig. s5. The task of the control system is to log the information gathered by the sensors (including the cameras) and to periodically update the duty cycle of the PWM signal sent to the motor allowing the automatic azimuth change of the torsion balance. The real-time data acquisition of the sensors (excluding the cameras) and the calculation of the feedback are realized by a microcontroller based embedded device. The cameras feed the images directly to a PC where they are saved and then processed. The image processing is executed real-time.

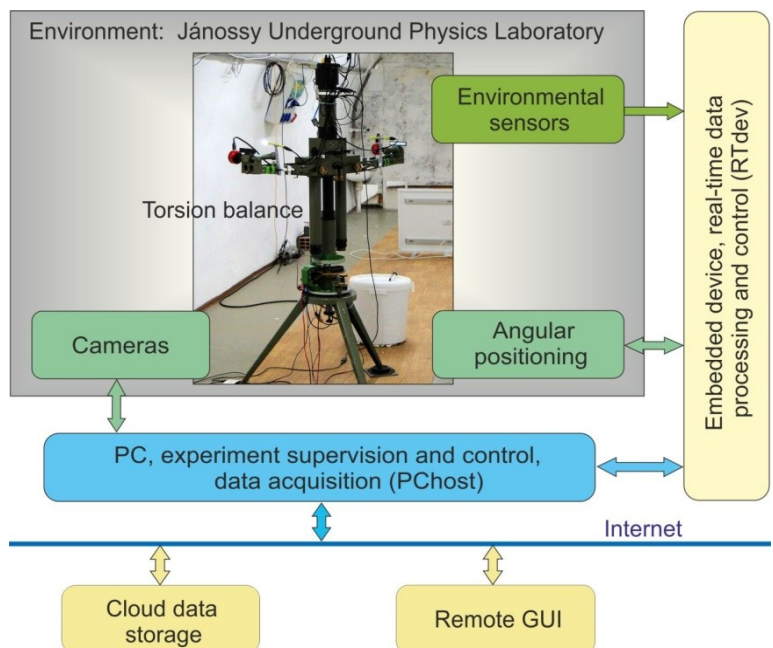

***Fig. s5***. *The control system built around the torsion balance*

Among the above-mentioned technical developments, the camera readout and the image processing are the crucial parts of the sensitivity improvement. The determination of the pendulum directions is based on the Fourier spectra of the visible part of the scale, with about altogether 40 lines. Therefore, the accuracy of the readout is below the resolution of the HD camera. Assuming independent scale divisions, that is a perfect scale, and considering that one pixel = 0.02 div, one can get the above mentioned 0.003 div sensitivity of the instrument.